# Depth Profiling of Oxygen Migration in Ta/$HfO_2$ Stacks During Ionic Liquid Gating

Beatrice Bednarz,[1,‡,*] Martin Wortmann,[2,‡] Olga Kuschel,[1,3] Fabian Kammerbauer,[1] Mathias Kläui,[1] Andreas Hütten,[2] Joachim Wollschläger,[3] Gerhard Jakob,[1] and Timo Kuschel[1,2,*]

[1)] *Institute of Physics, Johannes Gutenberg University Mainz, Staudingerweg 7, 55128 Mainz, Germany*
[2)] *Faculty of Physics, Bielefeld University, Universitätsstraße 25, 33615 Bielefeld, Germany*
[3)] *Faculty of Physics, Osnabrück University, Barbarastraße 7, 49076 Osnabrück, Germany*

‡ Equal author contribution
* Correspondence: bbednarz@uni-mainz.de, tkuschel@uni-mainz.de

## ABSTRACT

Ionic liquid (IL) gating has emerged as a powerful tool to control the structural, electronic, optical, and magnetic properties of materials by driving ion motion at solid interfaces. In magneto-ionic systems, electric fields are used to move ions, typically oxygen, from a donor layer into an underlying magnetic metal. Although oxygen distribution is key to enabling precise and stable control in magneto-ionic systems, the spatial distribution and voltage-dependence of oxygen incorporation in such nanoscale stacks remain unknown. Here, we quantify oxygen depth profiles and oxide formation in Si/ $SiO_2$/ Ta (15)/ $HfO_2$ (t) films after IL gating as a function of the gate voltage and $HfO_2$ capping thickness (2 and 3 nm). X-ray reflectivity and X-ray photoelectron spectroscopy measurements revealed a threshold electric field of ≈ -2.8 MV/cm to initiate oxygen migration from $HfO_2$ into metallic Ta. The resulting $Ta_2O_5$ thickness increases linearly with gate voltage, reaching up to 4 nm at -3 V gating. Notably, the required electric field rises with oxide thickness, indicating a progressively growing barrier for thicker oxide films. The Ta/$Ta_2O_5$ interface remains atomically sharp for all gate voltages. This suggests that complete $Ta_2O_5$ layers form sequentially before further oxygen penetration, with no sign of deeper diffusion into bulk Ta. Thinner capping layers enhance oxidation, relevant for optimized stack design. Additionally, indium migration from the indium tin oxide electrode to the sample surface was observed, which should be considered for surface-sensitive applications. These insights advance design principles for magneto-ionic and nanoionic devices requiring precise interface engineering.

**Keywords**: Voltage control of magnetism, oxygen doping, X-ray photoelectron spectroscopy, X-ray reflectivity, depth profile

Ionic liquids (IL) are well-known for their unique combination of high electrochemical capacitance, usually low toxicity, negligible volatility, and broad thermal stability range.[1] Their use in ionic gating enables the application of extremely large electric fields (10–100 MV/cm),[2] far exceeding the dielectric breakdown limits of conventional $SiO_2$-based metal–oxide–semiconductor field-effect transistors (MOSFETs).[2] These strong electric fields allow for field-driven ion motion, enabling reversible tuning of electronic, magnetic, optical, and structural properties.[2–6] In this regard, ionic liquid gating complements solid-state gating,[7–9] which typically requires comparatively thick dielectric layers to ensure electric insulation. The electric double layer of an IL is only about 1 nm thick, which enables the large electric fields characteristic of IL gating.[2] Similarly to solid-state gating, ionic liquid gating has proven effective in controlling a wide range of magnetic interactions, including magnetic anisotropy[10,11], exchange bias[12], Dzyaloshinskii–Moriya interaction[13], and Ruderman–Kittel–Kasuya–Yosida coupling[14]. This expanding toolbox of electric-field control over magnetic states opens possible pathways to ultra-low-power data storage, neuromorphic hardware, and spin-based sensors.[10,15–19]

Among various mobile species, oxygen ions are particularly attractive for magneto-ionic control due to their reactivity with a broad range of materials and the availability of multiple oxygen donor layers. Despite the widespread use of voltage-induced oxygen migration in functional magnetic stacks, the spatial profile of oxygen incorporation remains unknown. This is critical because the penetration depth, concentration profile, and sharpness of the resulting oxide/metal interface influence bulk magnetic and electronic properties, through changes in oxidation state or chemical composition.[10] The oxygen depth profile might also affect adjacent functional layers in ultrathin or multilayer architectures. For instance, in exchange bias systems or synthetic antiferromagnets, magnetic properties are set at buried interfaces. Information on the oxygen penetration depth is essential for determining the optimal thickness of layers above the critical interface to either modulate or protect the interface. Additionally, recent work has shown that introducing a thin tantalum (Ta) insertion layer below the IL can significantly enhance the cyclability of oxide-based magneto-ionic devices,[20] suggesting that the oxygen incorporation profile in such layers may play a key role in device stability and performance. However, it is neither known which amount of oxygen ions penetrates far down into the material and how the oxygen profile looks after IL gating nor which process underlies the migration. A more detailed understanding of the oxygen distribution is therefore key to enabling precise and stable control in magneto-ionic systems.

In this study, we combine X-ray reflectivity (XRR) with angle and energy resolved X-ray photoemission spectroscopy (XPS) to study the oxygen depth distribution in Ta as a function of the IL gate voltage in two systems with different hafnium oxide ($HfO_2$) capping layer thicknesses (see Fig. 1 a). Ta was chosen as the investigated metal because of its very smooth growth, making it an ideal model system for the investigation of homogeneous metal films with a thickness exceeding the maximum penetration depth of XPS. This allows for a clean signal without any influence of the substrate layer. Because of its very smooth growth, ductility and corrosion resistance, Ta is both a common seed layer as well as an important coating in a variety of applications.[21,22] $HfO_2$ was selected as a capping layer due to its high-$\kappa$ dielectric properties and high oxygen ion mobility, making it a common component in IL gating stacks.[23] Oxygen ions were driven into the Ta layer by applying gate voltages between -1.0 and -3.0 V using IL gating.

We demonstrate that the average oxide layer thickness increases linearly with gate voltage when voltages above -1.0 V are applied. For thinner $HfO_2$ capping layers, the average oxide thickness at the same applied gate voltage is larger. The oxide/metal interface thereby remains atomically sharp, only a few Å wide, at all gate voltages. Notably, indium ions from the indium tin oxide (ITO) electrode above the IL migrate to the sample surface during IL gating. This is important to consider when designing surface-sensitive applications.

## RESULTS AND DISCUSSION

Oxygen migration in magneto-ionic systems is investigated by XRR and XPS. As a model system, we studied Si/ $SiO_2$/ Ta (15)/ $HfO_2$ ($t$) heterostructures. Here, the number in the parentheses denotes layer thickness in nm and $t$ = 2 or 3 nm (Fig. 1 a). To resolve the oxygen distribution within the Ta layer, we first carried out XRR measurements (Fig. 1 b). The results for the samples with a 2 nm $HfO_2$ capping layer are shown in Figure 2. The XRR fits and corresponding scattering length density (SLD) graphs are exemplarily shown for the as-deposited state (Fig. 2 a) and after 10 minutes of maximum gate voltage of -3 V (Fig. 2 b). All other fits can be found in the supporting information (SI). For the fit model, we considered the $HfO_2$ capping layer, an oxidized $Ta_2O_5$ layer, which is the only stable oxide of Ta, the metallic Ta layer, a thin bottom $Ta_2O_5$ layer between the Ta and the $SiO_2$, as well as the $SiO_2$ substrate.

The SLD profiles of the samples depict a clear and only a few Å wide interface between the Ta and the upper, gating induced $Ta_2O_5$. However, the two oxidized layers above the Ta (in the applied fit model $Ta_2O_5$ and $HfO_2$) cannot be clearly distinguished as can be seen from the very broad transition between the two in the inset of Figure 2 b. The reason is the very similar density and therefore similar SLD of

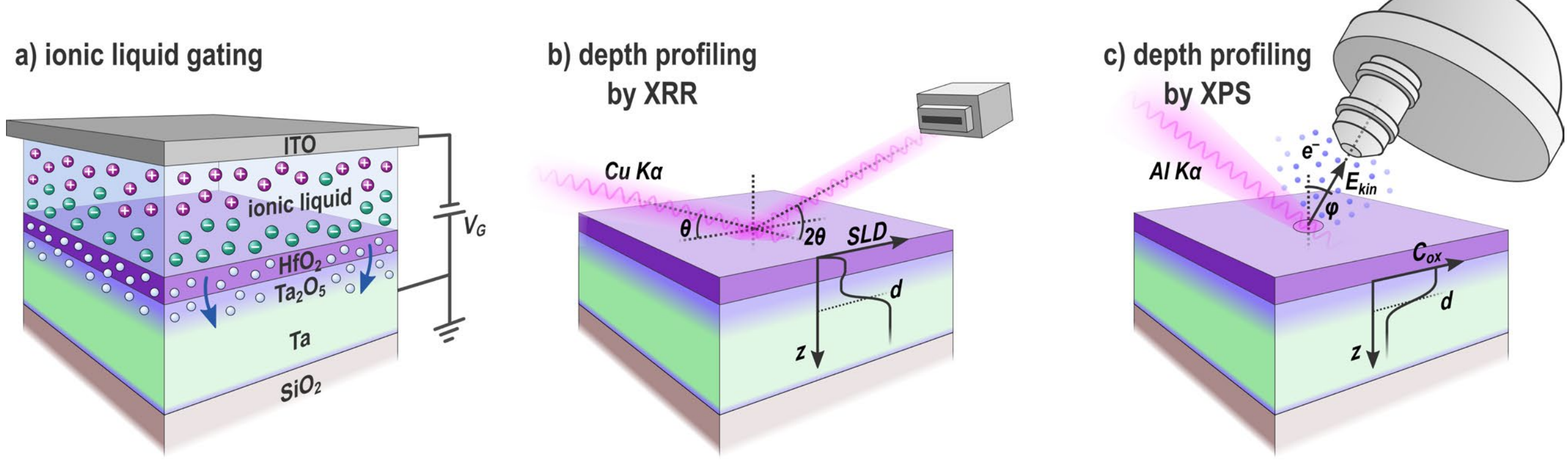

*Figure 1: (a) Sketch of the investigated material stack inside IL gating setup. By applying a negative gate voltage ($V_G$) the IL is polarized and transmits the gate voltage to the sample. There, oxygen ions (white spheres) are pushed from the $HfO_2$ into the Ta layer, forming a $Ta_2O_5$ layer at the interface. (b) Sketch of the XRR setup. By shining X-rays at grazing angles onto the sample and measuring their reflection in a specular geometry, information on the average layer thickness and interfacial roughness of the different layers can be obtained. The quantity which defines the scattering is the scattering length density (SLD, see SI7 for more details). (c) Sketch of the XPS setup, in which the photoelectrons are detected as a function of their emission angle $\varphi$. From their kinetic energy $E_{kin}$ the binding energy of the respective element can be obtained, which provides information about the element and its oxidation state. By varying the investigated emission angle, the depth distribution from which the photoelectrons stem changes. Using this, a depth profile of the concentration of oxidized Ta atoms ($C_{ox}$) can be deduced.*

$Ta_2O_5$ ($\rho$ = 8.2 g/cm$^3$ ≙ SLD = 1.95 $r_e/\text{Å}^3$) and $HfO_2$ ($\rho$ = 9.7 g/cm$^3$ ≙ SLD = 2.27 $r_e/\text{Å}^3$). For the same reason, additional intermediate $TaO_x$ compositions cannot be excluded, if their densities lie close enough to the densities of $Ta_2O_5$ and $HfO_2$. However, as shown by the XPS analysis, there are no signs of additional intermediate oxides. To disentangle the $Ta_2O_5$ and $HfO_2$ layers, the $HfO_2$ thickness was fixed to its value in the as-deposited state for all further gating steps. This is a reasonable assumption as the $HfO_2$ layer shows no sign of an induced metallic phase after gating, as demonstrated by XPS (Fig. 3). Similarly, the thickness and roughness of the bottom $Ta_2O_5$ was fixed to the value obtained from the as-deposited state, assuming that the applied field is screened by the Ta layer and does not influence the bottom $Ta_2O_5$ layer. In the following, $Ta_2O_5$ will therefore always refer to the upper $Ta_2O_5$ layer induced by the IL gating.

Figure 2 c shows the change of thickness of the upper $Ta_2O_5$ layer, the metallic Ta, as well as the total film thickness as a function of the applied gate voltage. A threshold voltage of approximately -1.0 V is required to oxidize Ta. Above this voltage, the average $Ta_2O_5$ thickness increases linearly with the applied gate voltage up to (3.9$\pm$0.3) nm after -3.0 V gating. In parallel, the Ta thickness reduces from 14.5 to 13.1 nm ($\pm$0.1 nm). This is notably less compared to the increase in the $Ta_2O_5$ thickness. The reason is the significantly lower mass density of $Ta_2O_5$ compared to metallic Ta. The total thickness therefore increases.

Figure 2 d presents the change in roughness at the upper $Ta_2O_5$/Ta interface with increasing gate voltage. In general, the interface roughness is very low, especially when compared to the layer thickness. The data suggests that the roughness increases with the initial gating and then slowly decreases again with increasing gate voltage. This increase likely originates from the structural reorganization of the topmost Ta layer and the density change accompanying the formation of the first $Ta_2O_5$ layer. The subsequent decrease in interface roughness after completion of the first layer indicates smooth growth of the additional $Ta_2O_5$ layers. Overall, the interface roughness only changes by slightly more than 0.1 nm ($\pm$0.05 nm), which is less than the thickness of a monolayer (0.17 nm for a perfect bcc Ta crystal with a lattice constant of 0.33 nm[24]).

To obtain more information about the chemical processes during IL gating and to validate the results from XRR, we performed XPS measurements. First of all, the nonvolatility of the oxidation was confirmed by repeating the XPS measurement of the sample with 2 nm $HfO_2$ gated at -2.0 V after 5 months. By analyzing the spectra from different emission angles (angle-resolved XPS) as well as peak regions (multiple-energies approach), information on the depth distribution of the elements can be obtained (see Fig. 1 c and reference [25]). Therefore, we fitted all four available Ta peak regions, Ta 4f, Ta 5s, Ta 4d and Ta 4p at six different emission angles, 0°, 15°, 25°, 35°, 45°, and 55° relative to the sample normal. For larger angles, the error on the effective attenuation length (EAL), which determines the probability that a photoelectron reaches the sample surface without scattering, becomes significantly larger.[26] Therefore, no angles larger than 55° were considered. The results from XRR were used as starting parameters for the fitting of all XPS peaks. More details on the fit parameters and procedure can be found in the SI.

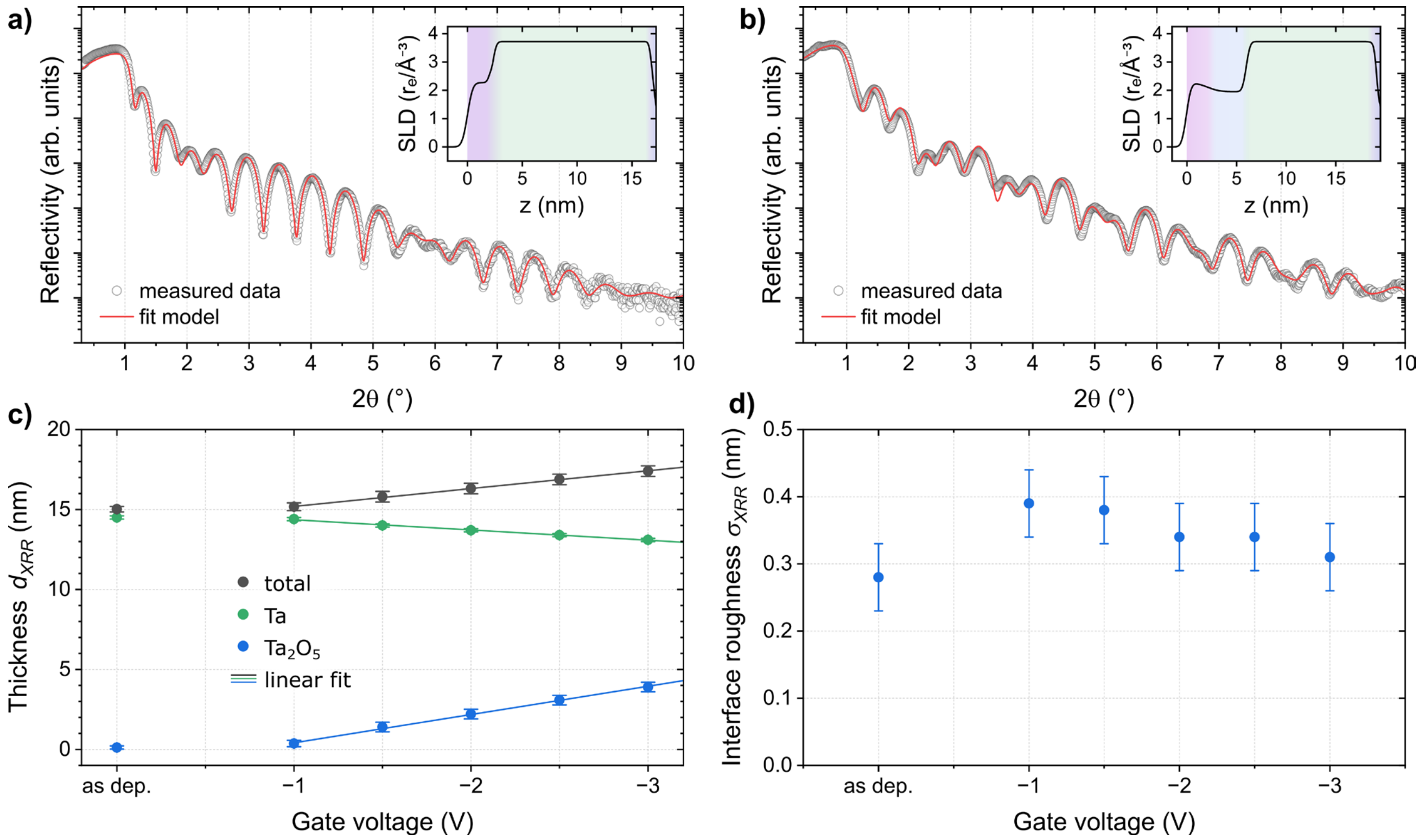

*Figure 2: XRR analysis of the sample Si/ $SiO_2$/ Ta (15)/ $HfO_2$ (2) as a function of the applied gate voltage. (a), (b) XRR fits for the samples in the as-deposited state and after applying -3.0 V for 10 min. The insets show the corresponding SLD profiles (starting from air with SLD = 0, to $HfO_2$ marked in purple, $Ta_2O_5$ in blue, Ta in green and the bottom $Ta_2O_5$ due to the interface with $SiO_2$ in blue again). (c) Evolution of the thicknesses of Ta, $Ta_2O_5$ and the total thickness (sum of Ta, $Ta_2O_5$ and the $Ta_2O_5$ below the Ta) as a function of the gate voltage, applied for 10 minutes each. (d) Roughness σ at the upper $Ta_2O_5$/Ta-interface. The XRR fits and SLD profiles at the other gate voltages, as well as details on the fits and all estimated uncertainties can be found in the SI.*

Figure 3 shows a subset of the fitted data for the sample with 2 nm $HfO_2$ capping. To investigate the effect of gating, the spectrum of the as-deposited state (Fig. 3 a) is compared to the state after applying the maximum gate voltage of -3.0 V for 10 minutes (Fig. 3 b). The survey spectra in Figures 3 a1 and b1 show the full accessible energy range. The three different energy regions with the clearest distinction between the metallic Ta ($Ta^0$, green) and oxidized $Ta_2O_5$ ($Ta^{5+}$, blue) peaks are shown in the close-ups below the surveys (Figs. 3 a2–4 and b2–4). In the as-deposited state, only the metallic $Ta^0$ and $Hf^{4+}$ ($HfO_2$, purple) show pronounced peaks. Almost no oxidized $Ta^{5+}$ is present. In contrast, after gating at -3.0 V for 10 min, the $Ta^0$ peaks get significantly smaller while $Ta^{5+}$ now shows large, distinct peaks, confirming a significant oxidation of the Ta layer. Notably, no additional peaks from intermediate Ta oxidation state are visible. Such peaks would be expected between the $Ta^0$ and $Ta^{5+}$ peaks, as was for example observed at the interface between Si and $SiO_2$.[27] As none of the gating steps shows such peaks (see SI), we conclude that an atomically sharp interface from Ta to $Ta_2O_5$ forms and propagates upon gating. This also matches the XRR results, which showed a sharp interface at all gate voltages.

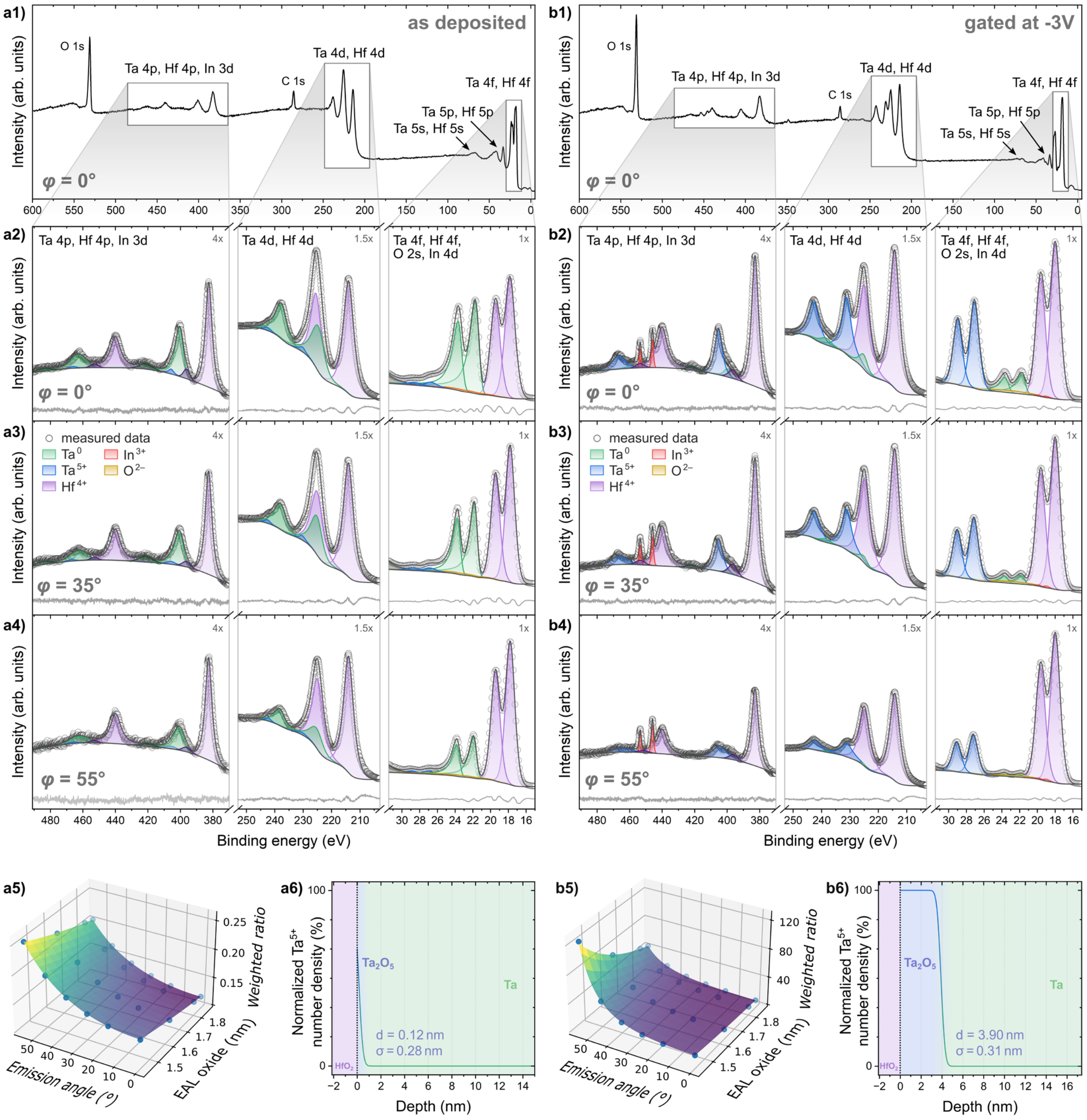

*Figure 3: XPS spectra for the sample with 2 nm $HfO_2$ (a) in the as-deposited state and (b) after the maximum applied gate voltage of -3.0 V. (a1, b1) Survey spectra and (a2-4, b2-4) fitted elemental spectra for the peak regions Ta 4p, 4d, and 4f for three different emission angles (0°, 35°, and 55° respective to the surface normal, for which cos(35°) is approximately in the middle between the cosines of the other two angles). The fitted elemental spectra for the regions Ta 5s and O 1s, as well as the other angles and gating steps, can be found in the SI. The Ta 5p region (starting at 31 eV towards higher binding energies) was not fitted because of the high number of peaks with large overlap and the resulting high uncertainty of the fitting. However, since this region lies right next to the Ta 4f peak region, a dummy peak (grey) was fitted into the first peak of the Ta 5p region to account for the residual signal. (a5, b5) Weighted $Ta^{5+}$:$Ta^{0}$ peak ratios for all angles and peak regions, fitted by eq. 1. Since the EALs for Ta and $Ta_2O_5$ are interdependent and defined by the kinetic energy of the photoelectrons, only the EAL of $Ta_2O_5$ is representatively plotted on the axis. (a6, b6) $Ta^{5+}$ depth profiles, corresponding to $d$ and $\sigma$ obtained from the fit in (a5, b5), respectively. The $Ta^{5+}$ number densities were normalized to their respective nominal maximum values.*

When looking at the $HfO_2$ peaks, it is noteworthy that no metallic $Hf^0$ peaks appear after gating. This is remarkable, as the oxygen causing the oxidation of Ta is moved down by the gate voltage from $HfO_2$ into the Ta layer. This can be explained by two possible scenarios: One explanation would be that possible oxygen voids in $HfO_2$ were filled again after the gating during the transfer to the XPS (approximately 1 day of transport, partly in air). However, breaking up $HfO_2$ bonds to access oxygen seems energetically unlikely. Therefore, the suggestion by Gilbert *et al.* on the basis of the enthalpies of formation, that the oxygen contributing to the gating could be interstitial oxygen, seems more reasonable.[28] In the case of IL gating, there might be an additional supply of oxygen through the IL.

Another noteworthy change after gating is the appearance of sharp indium peaks ($In^{3+}$, red) in the spectrum of Ta 4p and Hf 4p (Fig. 3 b2–4). Since this In is not present in the as-deposited state, it has to originate from the ITO coating of the glass plate, which is used as the top gate contact. Apparently, some of this In is released from the ITO during gating. The standard electrode potential of $In^{3+}$ to $In^0$ is -0.34 V vs. a standard hydrogen electrode, which is lower than the applied voltage. Therefore, likely In is reduced to $In^0$, and deposits on the sample surface. After gating, it oxidizes back to $In^{3+}$. The intensity of the In peaks does not significantly change for different emission angles, in contrast to the Ta peaks which become significantly smaller for larger angles. Even the adjacent Hf peak decreases slightly in size for increasing angles in comparison to the In peaks. Hence, the In atoms were deposited only at the surface of the sample and did not penetrate into the $HfO_2$.

Based on the finding that there are no intermediate oxidation states at the interface between Ta and $Ta_2O_5$, we can model the interface assuming a symmetric transition between Ta and $Ta_2O_5$ of the form of an error function. This is the common assumption for metallic interfaces, as also applied in XRR. The thickness $d$ of the $Ta_2O_5$ layer and roughness $\sigma$ of the interface between Ta and $Ta_2O_5$ can then be calculated using equation (1).

$$\frac{I_{\mathrm{Ta}^{5+}}\cdot N_{\mathrm{Ta}^0}}{I_{\mathrm{Ta}^0}\cdot N_{\mathrm{Ta}^{5+}}}\left(L_{\mathrm{Ta_2O_5}}, L_{\mathrm{Ta}}, \varphi\right) = \frac{L_{\mathrm{Ta_2O_5}}}{L_{\mathrm{Ta}}}\cdot \exp\left(\frac{d}{L_{\mathrm{Ta_2O_5}}\cos\varphi}-\frac{d}{L_{\mathrm{Ta}}\cos\varphi}\right) \cdot \frac{\operatorname{erfc}\left(-\frac{d}{\sqrt{2}\,\sigma}\right)-\operatorname{erfc}\left(\frac{\sqrt{2}\,\sigma}{2L_{\mathrm{Ta_2O_5}}\cos\varphi}-\frac{d}{\sqrt{2}\,\sigma}\right)\cdot\exp\left(\frac{2\sigma^2}{\left(2L_{\mathrm{Ta_2O_5}}\cos\varphi\right)^2}-\frac{d}{L_{\mathrm{Ta_2O_5}}\cos\varphi}\right)}{\operatorname{erfc}\left(\frac{d}{\sqrt{2}\,\sigma}\right)+\operatorname{erfc}\left(\frac{\sqrt{2}\,\sigma}{2L_{\mathrm{Ta}}\cos\varphi}-\frac{d}{\sqrt{2}\,\sigma}\right)\cdot\exp\left(\frac{2\sigma^2}{(2L_{\mathrm{Ta}}\cos\varphi)^2}-\frac{d}{L_{\mathrm{Ta}}\cos\varphi}\right)} \quad (1)$$

Thereby, $\frac{I_{\mathrm{Ta}^{5+}}\cdot N_{\mathrm{Ta}^0}}{I_{\mathrm{Ta}^0}\cdot N_{\mathrm{Ta}^{5+}}}$ denotes the normalized intensity ratio of the metallic and corresponding oxidized Ta peak inside the same peak region. The peak intensities $I_{\mathrm{Ta}^{5+}}$ and $I_{\mathrm{Ta}^0}$ of the oxidized and metallic Ta peaks are normalized to the atomic number densities $N_{\mathrm{Ta}^{5+}}$ and $N_{\mathrm{Ta}^0}$ of $Ta^{5+}$ and $Ta^0$ atoms per unit volume, respectively. The intensity ratio depends on the emission angle $\varphi$, as well as on the EALs $L_{\mathrm{Ta_2O_5}}$ and $L_{\mathrm{Ta}}$ in $Ta_2O_5$ and Ta, respectively, which are functions of the kinetic energy $E_{kin}$ of the photoelectrons of the respective peak region. More details on the equation and the required parameters can be found in the SI as well as in references [25] and [29].

For all peak regions, the normalized intensity ratios have to correspond to the same values of $d$ and $\sigma$. Therefore, all fits were optimized accordingly. Figures 3 a5 and b5 show the results of the weighted intensity ratios and the 3D fit of equation (1) as a function of $\varphi$ and $L_{\mathrm{Ta_2O_5}}(E_{\mathrm{kin}})$.

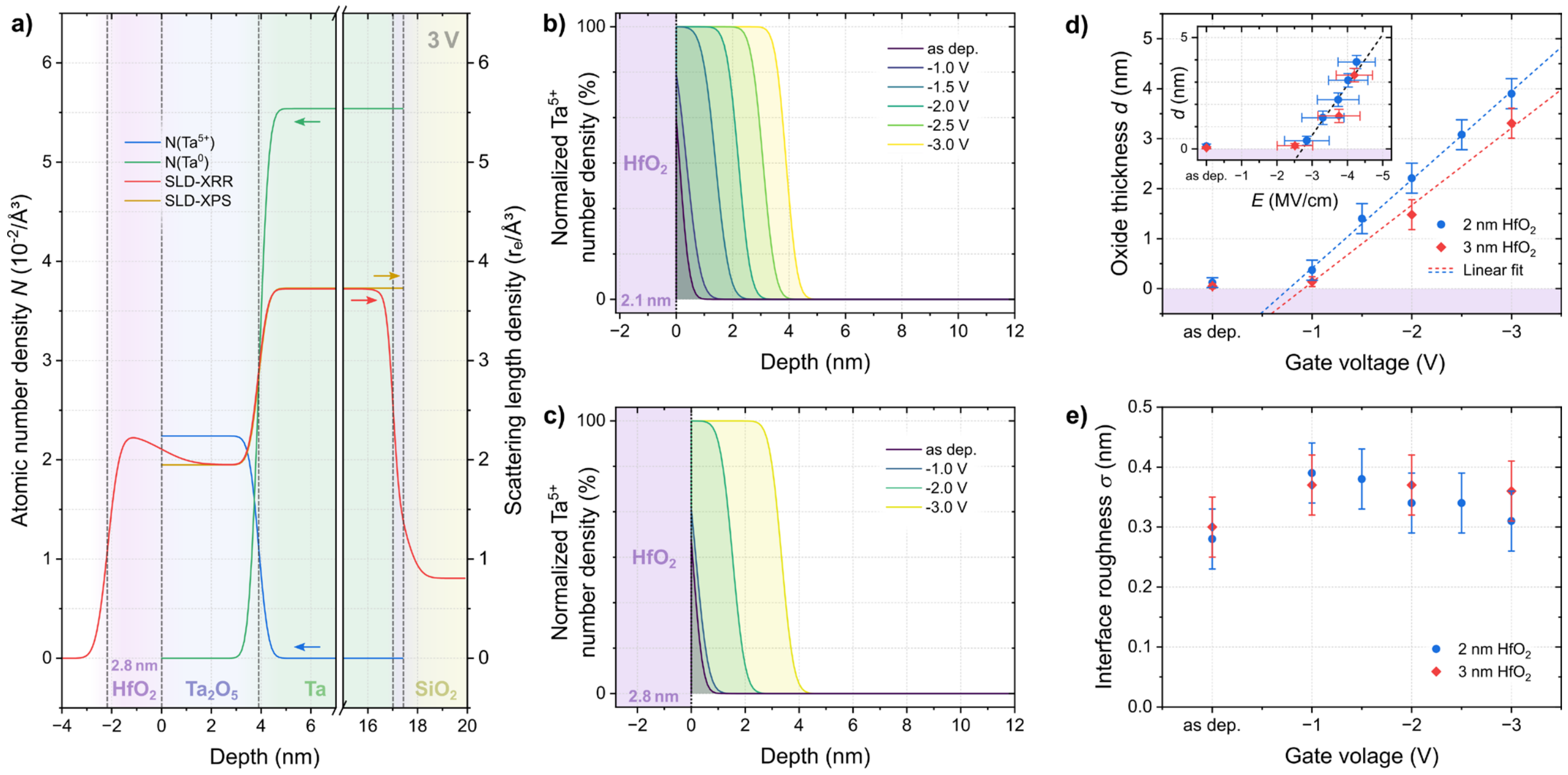

*Figure 4: (a) Comparison of the scattering length density (SLD) as obtained from XRR and the depth profiles as well as corresponding SLD (calculated for the energy of Cu Kα radiation to compare to XRR, details can be found in SI 7) as obtained from XPS for the sample with 2 nm $HfO_2$ capping and gated at -3.0 V. (b, c) $Ta^{5+}$ depth profiles for the samples with (b) 2 nm and (c) 3 nm nominal $HfO_2$ capping as a function of the gate voltage. (d) Average $Ta_2O_5$ layer thicknesses obtained from the depth profiles in (b) and (c) as a function of the gate voltage. The inset shows the same thicknesses as a function of the applied electric field (assuming an electric double layer of the IL of 1 nm, added up with the $HfO_2$ and $Ta_2O_5$ thicknesses). The errorbar gives the difference when a thickness of the electric double layer of 2 nm is assumed. (e) Interface roughness at the $Ta_2O_5$/Ta interface as a function of the gate voltage as obtained from the depth profiles in (b) and (c).*

From the corresponding results of the fit for $d$ and $\sigma$, we obtained the $Ta^0$ and $Ta^{5+}$ depth profiles (Figs. 3 a6 and b6). Both $d$ and $\sigma$ were thereby consistent with the results from XRR, such that the values obtained from both methods are identical.

Figure 4 a shows the comparison between the SLD profile obtained from XRR and the depth profiles obtained from XPS, given by the number densities $N(Ta^0)$ and $N(Ta^{5+})$, after gating at -3.0 V. When normalized and expressed in percent, $N(Ta^{5+})$ is identical to the $Ta^{5+}$ depth profile, as plotted in Figures 4 b and c. For a more direct comparison between the results from XRR and XPS, the SLD-XPS profile was calculated from $N(Ta^0)$ and $N(Ta^{5+})$ at the energy of Cu K$\alpha$ radiation. Details on the calculation can be found in SI 7. In XPS, we only investigated the interface between Ta and $Ta_2O_5$, so the SLD-XPS was only calculated for that interface. The SLD profiles obtained from XPS and XRR are almost perfectly congruent at that interface. This confirms that both methods model the interface in the same way, using an error function, and yield the same average position and width of the interface. To focus on the $Ta_2O_5$/Ta interface, we will use the $Ta^{5+}$ depth profiles from XPS in the following to discuss the evolution of interface position and width upon gating, as well as the influence of the $HfO_2$ thickness.

The $Ta^{5+}$ depth profiles are depicted in Figure 4 b and c as a function of the gate voltage, for two different $HfO_2$ thicknesses. The results are qualitatively the same for both $HfO_2$ thicknesses, showing a sharp interface between Ta and $Ta_2O_5$, which is pushed deeper into the Ta layer with increasing gate voltage. Figure 4 d shows the comparison of the corresponding average oxide layer thicknesses $d$ for all samples. For both $HfO_2$ thicknesses, the threshold voltage required to significantly oxidize the Ta is around -1.0 V. Above this voltage, the average oxide layer thickness increases roughly linearly with the applied gate voltage in the investigated range. Overall, the oxide layer thickness is larger for all gate voltages when a thinner $HfO_2$ capping of only 2 nm is used. This is most likely due to the larger electric field created by applying an identical gate voltage over a thinner dielectric material.

In the inset (Fig. 4 d), the average oxide layer thickness is therefore plotted as a function of the electric field, calculated by assuming an IL electric double layer thickness of 1 nm. This value corresponds approximately to the combined thickness of one negatively and one positively charged ion layer, across which the voltage drop in the IL occurs. The relation between $d$ and $E$ is very similar for both $HfO_2$ thicknesses, with the same initial threshold electric field of $\approx$ -2.8 MV/cm to start oxidizing the Ta. Remarkably, the required electric field increases by $\approx$ -0.45 MV/cm for every additional nm of $Ta_2O_5$. This shows that the threshold electric field needed to drive oxygen ions from $HfO_2$ to the $Ta_2O_5$/Ta interface is not constant – as might be expected – but grows with oxide thickness. One possible contribution is the self-passivating nature of Ta, for which the energetic cost of further oxidation increases with depth.[29] This is also consistent with temperature-dependent oxide growth reported in the literature.[30] Additional factors, such as incomplete saturation of $Ta_2O_5$ during the 10-minutes gating period or a voltage-dependent increase in the IL's electric double-layer thickness, could also contribute, though these effects are likely too small to account for the observed slope.

The roughness of the interface shows a very similar behavior independently of the $HfO_2$ thicknesses (Fig. 4 e). Also, for the samples with thicker $HfO_2$, the roughness increases slightly, by approximately $(0.07\pm0.05)$ nm, when the initial gate voltage is applied. For increasing gate voltages, it shows a slight decrease, which is however less pronounced compared to the samples with thinner $HfO_2$ capping layer, and cannot be confirmed within the error margin.

**CONCLUSION**

Our results demonstrate that oxygen ions can be driven several nm into metallic films. In Ta, penetration depths up to 4 nm were achieved by applying -3 V for 10 minutes. This depth exceeds the typical thickness of magnetic thin films, which are usually 0.8–0.9 nm for out-of-plane magnetized layers and 1.0–2.5 nm for in-plane layers. Spacer layers, such as in synthetic antiferromagnets or magnetic tunnel junctions, can be even thinner. Thus, the observed oxygen penetration depth is sufficient to alter the properties of buried layers or interfaces. Depending of the application, this effect can be exploited deliberately or must be carefully managed when preservation of lower interfaces is required.

Furthermore, we reveal that the $Ta_2O_5$/Ta interface propagates linearly with applied voltage into the Ta film. The interface remains atomically sharp (few Å) at all voltages, indicating that complete $Ta_2O_5$ layers form sequentially. This behavior is consistent with electric-field screening. The sharpness of the interface further shows that oxidation occurs uniformly across the analyzed sample region, which reflects the high uniformity of the sputtered Ta film. In other materials, additional effects such as grain boundary diffusion may play a role. For example, Gilbert *et al.* reported that solid-state gating in Pd/Co/$AlO_x$/$GdO_x$ stacks leads to two distinct magnetic phases.[28] They attributed this to preferential oxygen transport along grain boundaries combined with slower diffusion within grains. In such cases, the process inside each grain or homogeneous region will likely resemble the behavior identified in this study. Thus, our findings provide a framework to interpret oxide depth profiles and disentangle competing physical effects in a broad range of systems.

When choosing the thickness of the dielectric capping layer, it is important to note that thinner $HfO_2$ enhances oxidation by increasing the electric field. Still, for reliable IL gating, a continuous capping layer is essential. The minimum thickness is therefore limited by the roughness of the underlying interface and growth characteristics of the capping layer.

Finally, we observe the migration of indium atoms from the ITO electrode onto the sample surface during gating. This is important to consider for all applications requiring a well-defined surface, *e.g.* when post-processing the sample after removal of the IL and gate.

The results from XRR and XPS showed very good agreement, thereby validating the derived depth profiles. In the present case, XRR provided depth profiles with lower uncertainty while requiring less experimental effort. However, the uncertainty of the extracted depth profiles is material-dependent, and the relative advantages of the two techniques may shift for different systems. Specifically, XRR yields more reliable results in cases of pronounced electron density contrasts, while XPS achieves higher sensitivity when several well-separated emission peaks with large chemical shifts are present. Beyond depth profiling, the evaluation of the full XPS spectrum using the multiple-energies approach provided valuable insights into oxidation states and enabled the detection of indium on the sample surface. Hence, the two techniques offer complementary strengths for a more robust assessment of interfacial structure and composition.

The presented results elucidate the mechanism of oxygen migration and $Ta_2O_5$ formation during IL gating, providing quantitative design rules for magneto-ionic stacks. This understanding is key for advancing low-power nanoelectronic, spintronic and neuromorphic devices.

## EXPERIMENTAL SECTION

**Sample growth.** The samples were grown on p-doped thermally oxidized $Si/SiO_2$ at room temperature in an industrial Singulus Rotaris magnetron sputtering tool. In this way, a very high sample quality and homogeneity was achieved, as evidenced by the smooth growth. All samples with the same nominal $HfO_2$ thickness were grown together on the same wafer and afterwards cut into 5 x 8 $cm^2$ large pieces.

**IL gating.** The IL 1-ethyl-3-methylimidazolium-bis-(trifluoromethylsulfonyl)-imide ($[EMIM]^+$ $[TFSI]^-$) (CAS: 174899-82-2, SKU: 711691-100G, >98 % (HPLC), from Sigma-Aldrich) was used to apply the gate voltages. A drop of the IL was placed in each corner of the $HfO_2$ surface. A glass slide (area $5 \times 7$ $mm^2$) coated with ITO, floating on top of the IL, was used as the top contact for the gate voltage. To ground the Ta, serving as the bottom contact, the Ta was wirebonded through the insulating layers on top. By applying negative gate voltages to this setup, oxygen ion species can be moved from $HfO_2$ into the Ta. Gate voltages between -1.0 and -3.0 V were applied at room temperature using a Keithley 2400. Thereby, the voltage was slowly increased, to minimize instabilities, up to the target voltage which was applied for 10 minutes. Afterwards, the IL was washed off in acetone.

**XRR measurements.** XRR measurements were performed using a Bruker D8 X-ray diffractometer using Cu Kα radiation. The analysis was done using the software GenX version 3.7.4. All data was measured in a specular geometry. Therefore, the spec_nx mode of the advanced reflectivity plugin was used, which fits the data using the Parrat algorithm.[31] The roughness is defined as the root mean square roughness at the top interface of the layer, assuming a Gaussian deviation from the ideal interface.

**XPS measurements.** XPS measurements were performed using an ESCA-unit Phi 5000 VersaProbe III photoelectron spectrometer. Monochromated Al Kα radiation (1486.6 eV) with 25 W at 15 kV was used as the X-ray source. The resulting analyzed area has a diameter of 100 µm. The electron analyzer has a work function of 4.39 eV and was positioned at a fixed angle of 45° towards the X-ray source. For each sample, a survey (pass energy 224 eV) as well as the O 1s and Ta 4p, 4d, 5s and 4f energy ranges (pass energy 55 eV) were recorded for electron emission angles $\varphi$ between 0° and 55° relative to the surface normal by tilting the sample. The analysis was done using the software CasaXPS version 2.3.26PR1.0.

## ASSOCIATED CONTENT

**Supporting Information (SI) Available:** Additional Figures for XRR and XPS measurements, tables with the used material constants, list of peak positions and parameters for XPS peak fits and further information on the equations used for the analysis.

## ACKNOWLEDGEMENTS

This project has received funding from the European Union's Horizon 2020 Research and Innovation Programme under the Marie Skłodowska-Curie grant agreement No 860060 "Magnetism and the effect of Electric Field" (MagnEFi), as well as from the Deutsche Forschungsgemeinschaft (DFG, German Research Foundation) – TRR 173/2-#268565370 Spin+X (Projects A01 and B02).

## AUTHOR DECLARATIONS

**Conflict of interest**
The authors have no conflicts to disclose.

## DATA AVAILABILITY

Data that support the findings of this study will be openly available in Zenodo upon acceptance of this manuscript.

# Supporting Information (SI)

## Depth Profiling of Oxygen Migration in Ta/$HfO_2$ Stacks During Ionic Liquid Gating

Beatrice Bednarz,[1,‡,*] Martin Wortmann,[2,‡] Olga Kuschel,[1,3] Fabian Kammerbauer,[1] Mathias Kläui,[1] Andreas Hütten,[2] Joachim Wollschläger,[3] Gerhard Jakob,[1] and Timo Kuschel[1,2,*]

[1)] *Institute of Physics, Johannes Gutenberg University Mainz, Staudingerweg 7, 55128 Mainz, Germany*
[2)] *Faculty of Physics, Bielefeld University, Universitätsstraße 25, 33615 Bielefeld, Germany*
[3)] *Faculty of Physics, Osnabrück University, Barbarastraße 7, 49076 Osnabrück, Germany*

‡ Equal author contribution
* Correspondence: bbednarz@uni-mainz.de, tkuschel@uni-mainz.de

## SI 1. X-ray reflectivity (XRR) fit parameters

The XRR analysis was performed in GenX[1] (homepage: https://aglavic.github.io/genx/) version 3.7.4, using the following slab model: $SiO_2$ / $Ta_2O_5$ bottom / Ta / $Ta_2O_5$ top / $HfO_2$.

For all layers, literature densities were used. For $HfO_2$ and the top $Ta_2O_5$, the reason is their very similar mass density and correspondingly similar scattering length density (SLD). Therefore, they are nearly indistinguishable in XRR, which leads to unreasonable density values when fitted freely. For Ta, a free fit gave densities very close to the literature density (within 0.3 g/cm$^3$) which however differed between the samples after the different gating steps. Since all samples with the same $HfO_2$ thickness were sputter deposited together on the same wafer, their Ta densities cannot be different. Therefore, the literature density was set for all samples. For the bottom $Ta_2O_5$ layer, the thickness is too thin to fit the density freely.

Additionally, for the $HfO_2$ layer and the bottom $Ta_2O_5$ layer, the thicknesses were fixed according to their thickness in the corresponding as-deposited sample. For the bottom $Ta_2O_5$ the underlying assumption is that the Ta layer above screens the electric field and therefore the bottom $Ta_2O_5$ layer is not significantly affected by the gating. Also the roughness of this layer was fixed according to the as-deposited state, using the knowledge that the roughness of the $SiO_2$ wafer is in the range of 1-3 Å. For $HfO_2$, the reason to fix the thickness is the indistinguishability between the $HfO_2$ and the top $Ta_2O_5$ layer because of their similar densities. Using the knowledge from XPS, that the $HfO_2$ does not get reduced to metallic Hf (no $Hf^0$ peak and also no shoulder or other sign of a change due to gating appears in XPS), the assumption that the $HfO_2$ layer remains unaffected by gating seems justified. Still, this uncertainty was included in the error on the $Ta_2O_5$ thickness (see error discussion below).

Table S*1* shows the parameters fixed in the XRR fit. Tables S*2* and S*3* provide the corresponding results for the layer thickness $d$ and roughness $\sigma$ for the samples with 2 and 3 nm $HfO_2$, respectively. Figure S*1* shows the corresponding fits and SLD profiles for all samples.

Tables S*2* and S*3* also give the estimated uncertainties for all values. For the layer thicknesses, the uncertainty of the fits is estimated to be 0.1 nm. Additionally, the $HfO_2$ layer has an increasing uncertainty due to the assumption that it is unaffected by gating. This error gets propagated to the thickness of the upper $HfO_2$ layer because of their similar densities. For the roughness, the basic fitting uncertainty is estimated to be 0.05 nm. However, towards the sample surface (for $HfO_2$) and the substrate ($Ta_2O_5$ bottom) the uncertainty is larger and estimated to be 0.1 nm. At the $HfO_2$/$Ta_2O_5$ interface, the error is significantly larger because of the similar densities.

**Table S1:** Fixed parameters of XRR fits. Densities taken from literature values given by Merck KGaA and Kurt J. Lesker Company are used.

| | | $HfO_2$ | $Ta_2O_5$ top | Ta | $Ta_2O_5$ bottom | $SiO_2$ |
|---|---|---|---|---|---|---|
| Ta (15)/ $HfO_2$ (2) | $\rho$ (g/cm$^3$) | 9.7 | 8.2 | 16.7 | 8.2 | 2.7 |
| | $d$ (nm) | 2.1 | | | 0.4 | |
| | $\sigma$ (nm) | | | | 0.2 | |
| Ta (15)/ $HfO_2$ (3) | $\rho$ (g/cm$^3$) | 9.7 | 8.2 | 16.7 | 8.2 | 2.7 |
| | $d$ (nm) | 2.8 | | | 0.4 | |
| | $\sigma$ (nm) | | | | 0.3 | |

**Table S*2*:** Fitted parameters from XRR for the samples Si/ $SiO_2$/ Ta (15)/ $HfO_2$ (2) for all applied gate voltages. The values of the roughness $\sigma$ are defined at the upper interface of each layer. The thickness $d$ of the $Ta_2O_5$ layer and roughness $\sigma$ at the $Ta_2O_5$/Ta interface, which are particularly discussed in the manuscript, are highlighted with a green background. The grey values were fixed (see table S*1*) and are given here together with their estimated uncertainties. $SiO_2$ is not included because the substrate is assumed to have infinite thickness in XRR. The substrate roughness was fixed to the range 1–3 Å.

| | | $HfO_2$ | $Ta_2O_5$ top | Ta | $Ta_2O_5$ bottom |
|---|---|---|---|---|---|
| As deposited | $d$ (nm) | 2.1 ± 0.1 | 0.12 ± 0.10 | 14.5 ± 0.1 | 0.4 ± 0.1 |
| | $\sigma$ (nm) | 0.4 ± 0.1 | 0.12 ± 0.05 | 0.28 ± 0.05 | 0.2 ± 0.1 |
| -1.0 V | $d$ (nm) | 2.1 ± 0.2 | 0.4 ± 0.2 | 14.4 ± 0.1 | 0.4 ± 0.1 |
| | $\sigma$ (nm) | 0.4 ± 0.1 | 0.4 ± 0.2 | 0.39 ± 0.05 | 0.2 ± 0.1 |
| -1.5 V | $d$ (nm) | 2.1 ± 0.3 | 1.4 ± 0.3 | 14.0 ± 0.1 | 0.4 ± 0.1 |
| | $\sigma$ (nm) | 0.5 ± 0.1 | 0.5 ± 0.3 | 0.38 ± 0.05 | 0.2 ± 0.1 |
| -2.0 V | $d$ (nm) | 2.1 ± 0.3 | 2.2 ± 0.3 | 13.7 ± 0.1 | 0.4 ± 0.1 |
| | $\sigma$ (nm) | 0.5 ± 0.1 | 0.4 ± 0.3 | 0.34 ± 0.05 | 0.2 ± 0.1 |
| -2.5 V | $d$ (nm) | 2.1 ± 0.3 | 3.1 ± 0.3 | 13.4 ± 0.1 | 0.4 ± 0.1 |
| | $\sigma$ (nm) | 0.4 ± 0.1 | 0.6 ± 0.5 | 0.34 ± 0.05 | 0.2 ± 0.1 |
| -3.0 V | $d$ (nm) | 2.1 ± 0.3 | 3.9 ± 0.3 | 13.1 ± 0.1 | 0.4 ± 0.1 |
| | $\sigma$ (nm) | 0.4 ± 0.1 | 1.0 ± 0.7 | 0.31 ± 0.05 | 0.2 ± 0.1 |

**Table S*3*:** Fitted parameters from XRR for the samples Si/ $SiO_2$/ Ta (15)/ $HfO_2$ (3) at all applied gate voltages, presented equivalently to table S*2* for the samples with 2 nm $HfO_2$.

| | | $HfO_2$ | $Ta_2O_5$ top | Ta | $Ta_2O_5$ bottom |
|---|---|---|---|---|---|
| As deposited | $d$ (nm) | 2.8 ± 0.1 | 0.05 ± 0.05 | 14.5 ± 0.1 | 0.4 ± 0.1 |
| | $\sigma$ (nm) | 0.4 ± 0.1 | 0.07 ± 0.05 | 0.30 ± 0.05 | 0.3 ± 0.1 |
| -1.0 V | $d$ (nm) | 2.8 ± 0.2 | 0.14 ± 0.10 | 14.5 ± 0.1 | 0.4 ± 0.1 |
| | $\sigma$ (nm) | 0.4 ± 0.1 | 0.2 ± 0.1 | 0.37 ± 0.05 | 0.3 ± 0.1 |
| -2.0 V | $d$ (nm) | 2.8 ± 0.3 | 1.5 ± 0.3 | 14.0 ± 0.1 | 0.4 ± 0.1 |
| | $\sigma$ (nm) | 0.3 ± 0.1 | 0.4 ± 0.2 | 0.37 ± 0.05 | 0.3 ± 0.1 |
| -3.0 V | $d$ (nm) | 2.8 ± 0.3 | 3.3 ± 0.3 | 13.2 ± 0.1 | 0.4 ± 0.1 |
| | $\sigma$ (nm) | 0.4 ± 0.1 | 0.6 ± 0.4 | 0.36 ± 0.05 | 0.3 ± 0.1 |

## SI 2. Overview over all XRR fits

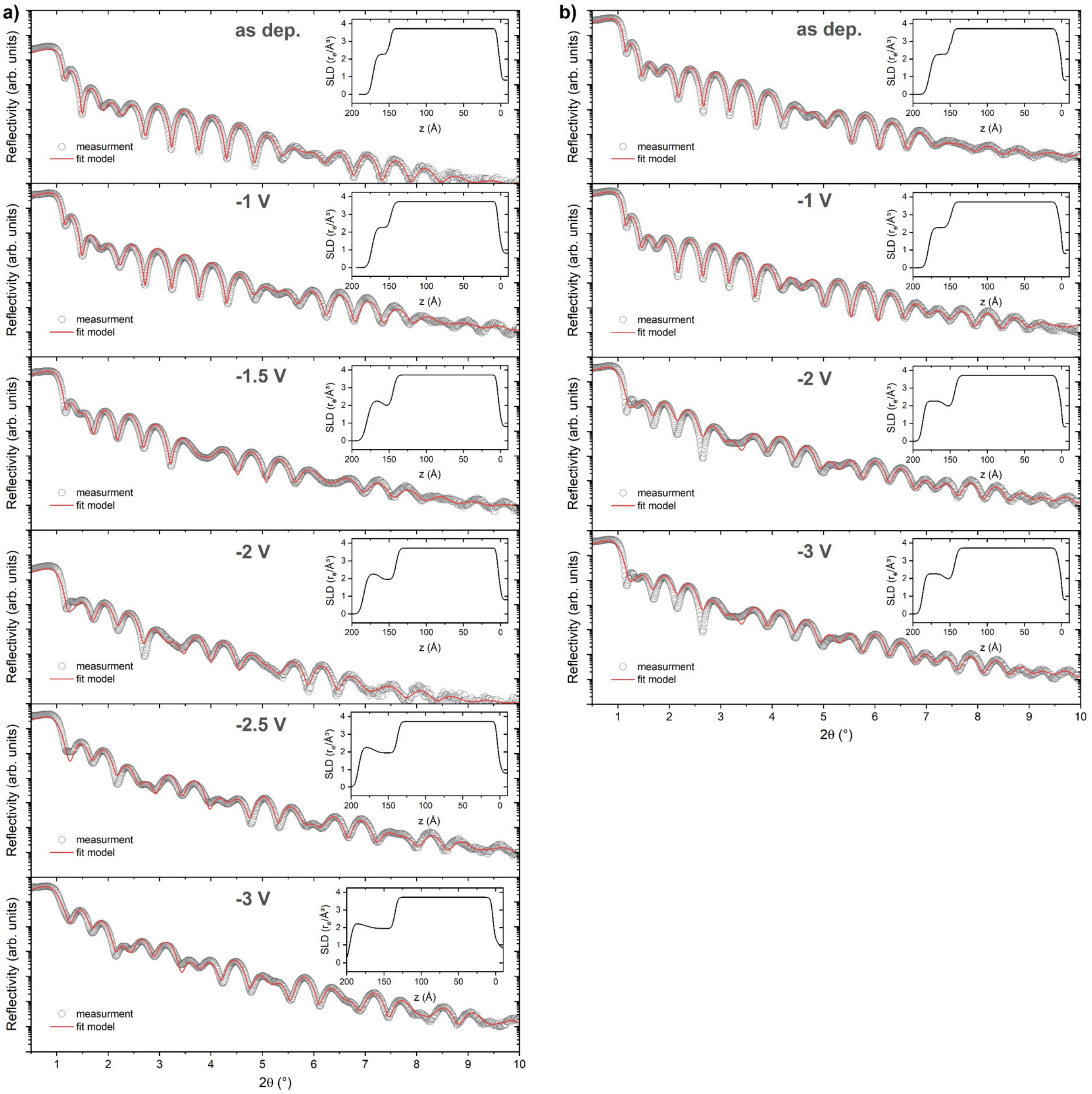

**Figure S*1*:** XRR fits and corresponding SLDs for the samples with (a) 2 and (b) 3 nm $HfO_2$, respectively, at all gate voltages. The narrow oscillation corresponds to the Ta thickness, while the wider oscillation with the visibly decreasing period corresponds to the combined thickness of $HfO_2$ and the top $Ta_2O_5$.

## SI 3. Origin of the equation used to obtain $d$ and $\sigma$ from the XPS data

In XPS, the binding energy of the emitted photoelectrons provides information on the corresponding element and its oxidation state. The number of photoelectrons $\Delta I$, which reach the surface without inelastic scattering, can be expressed according to the Beer-Lambert law, as

$$\Delta I = \Delta I_0 \, e^{\frac{-z}{L\cos(\varphi)}} \quad \text{with} \quad \Delta I_0 \propto \Delta z \tag{S1}$$

with the excited number of photoelectrons $\Delta I_0$ created in a thin layer with thickness $\Delta z$ at a depth $z$ below the surface, the distance $z/\cos(\varphi)$ travelled by the photoelectrons emitted at an angle $\varphi$ towards the sample normal and the inelastic mean free path (IMFP) or effective attenuation length (EAL) $L$ of the material (an explanation of the difference between the IMFP and EAL can be found in SI 4). In this study, we consider a two-layer system with a $Ta_2O_5$ layer with average thickness $d$ on top of a Ta film. We assume a gradual interface with a width $\sigma$ between the two layers which can be approximated by a complementary error function $\mathrm{erfc}\left(\frac{z-d}{\sqrt{2}\,\sigma}\right)$. In this case, the number of photoelectrons from the $Ta_2O_5$ layer $I_{\mathrm{Ta}^{5+}}$ and from the Ta layer $I_{\mathrm{Ta}^{0}}$ can be expressed as

$$\Delta I_{\mathrm{Ta}^{5+}} = \Delta I_{0,\mathrm{Ta}^{5+}} \exp\left(\frac{-z}{L_{\mathrm{Ta_2O_5}}\cos(\varphi)}\right) \tag{S2}$$

with $\Delta I_{0,\mathrm{Ta}^{5+}} \propto N_{\mathrm{Ta}^{5+}} \Delta z \, \mathrm{erfc}\left(\frac{z-d}{\sqrt{2}\,\sigma}\right)$

$$\Delta I_{\mathrm{Ta}^{0}} = \Delta I_{0,\mathrm{Ta}^{0}} \exp\left(\frac{-(z-d)}{L_{\mathrm{Ta}}\cos(\theta)}\right) \exp\left(\frac{-d}{L_{\mathrm{Ta_2O_5}}\cos(\varphi)}\right) \tag{S3}$$

with $\Delta I_{0,\mathrm{Ta}^{0}} \propto N_{\mathrm{Ta}^{0}} \Delta z \, \mathrm{erfc}\left(\frac{d-z}{\sqrt{2}\,\sigma}\right)$ .

Thereby, $\Delta I_{0,\mathrm{Ta}^{5+}}$ and $\Delta I_{0,\mathrm{Ta}^{0}}$ are the excited number of photoelectrons reaching the detector from a thin layer with thickness $\Delta z$ in the oxide and metal, respectively, $L_{\mathrm{Ta_2O_5}}$ and $L_{\mathrm{Ta}}$ are the EALs of the oxide and the metal and $N_{\mathrm{Ta}^{5+}}$ and $N_{\mathrm{Ta}^{0}}$ denote the number densities of oxidized or metallic Ta atoms, respectively. Integrating these two equations from $0 \leq z \leq \infty$ (assuming that the thickness of the Ta layer is much thicker than $L_{\mathrm{Ta}}$) yields the respective XPS peak intensities $I_{\mathrm{Ta}^{5+}}$ and $I_{\mathrm{Ta}^{0}}$. By calculating the quotient of the two and rearranging, we obtain equation 1 from the main text

$$\frac{I_{\mathrm{Ta}^{5+}} \cdot N_{\mathrm{Ta}^{0}}}{I_{\mathrm{Ta}^{0}} \cdot N_{\mathrm{Ta}^{5+}}}\left(L_{\mathrm{Ta_2O_5}}, L_{\mathrm{Ta}}, \varphi\right)$$
$$= \frac{L_{\mathrm{Ta_2O_5}}}{L_{\mathrm{Ta}}} \cdot \exp\left(\frac{d}{L_{\mathrm{Ta_2O_5}}\cos\varphi} - \frac{d}{L_{\mathrm{Ta}}\cos\varphi}\right) \cdot \frac{\mathrm{erfc}\left(-\frac{d}{\sqrt{2}\,\sigma}\right) - \mathrm{erfc}\left(\frac{\sqrt{2}\,\sigma}{2L_{\mathrm{Ta_2O_5}}\cos\varphi} - \frac{d}{\sqrt{2}\,\sigma}\right) \cdot \exp\left(\frac{2\sigma^2}{\left(2L_{\mathrm{Ta_2O_5}}\cos\varphi\right)^2} - \frac{d}{L_{\mathrm{Ta_2O_5}}\cos\varphi}\right)}{\mathrm{erfc}\left(\frac{d}{\sqrt{2}\,\sigma}\right) + \mathrm{erfc}\left(\frac{\sqrt{2}\,\sigma}{2L_{\mathrm{Ta}}\cos\varphi} - \frac{d}{\sqrt{2}\,\sigma}\right) \cdot \exp\left(\frac{2\sigma^2}{(2L_{\mathrm{Ta}}\cos\varphi)^2} - \frac{d}{L_{\mathrm{Ta}}\cos\varphi}\right)} \tag{S4}$$

Note that the peak region of each orbital corresponds to a different binding energy $E_{BE}$ and corresponding kinetic energy $E_{\mathrm{kin}}$ of the electrons, with $E_{\mathrm{kin}} = h\nu - E_{\mathrm{BE}} - \Phi$ with photon energy $h\nu = 1486.6$ eV and work function of the electron analyzer $\Phi = 4.39$ eV. This results in a unique set of EALs $L_{\mathrm{Ta_2O_5}}(E_{\mathrm{kin}})$ and $L_{\mathrm{Ta}}(E_{\mathrm{kin}})$. Therefore, this equation is a function of the emission angle $\theta$ as well as the kinetic energy $E_{\mathrm{kin}}$ of the electrons. The uncertainties on the oxide thickness $d$ and interface roughness $\sigma$, obtained from fitting this equation, can be minimized by increasing the number of analyzed peak regions and emission angles. A detailed error discussion and more information on this equation can be found in reference [2].

## SI 4. Calculation of the effective attenuation lengths (EALs) $L$ and number densities $N$

Equation (S4) (which is equal to equation (1) in the main text) requires the effective attenuation lengths (EALs) in $Ta_2O_5$ and Ta as an input parameter. The EAL is a measure of the attenuation of the photoelectrons inside the material, similar to the inelastic mean free path (IMFP). However, in contrast to the IMFP, the EAL also considers elastic scattering. Therefore, it is generally considered more reliable and was used in this analysis. To calculate the EALs $L_{\mathrm{Ta_2O_5}}$ and $L_{\mathrm{Ta}}$ inside $Ta_2O_5$ and Ta, respectively, the universal curve proposed by Seah was used[3]

$$L = \frac{(5.8+0.0041\cdot Z^{1.7}+0.088\cdot {E_{\mathrm{kin}}}^{0.93})\ a^{1.82}}{Z^{0.38}(1-0.02\cdot E_g)} \quad . \tag{S5}$$

Thereby, $E_{\mathrm{kin}}$ denotes the kinetic energy of the photoelectrons and $E_g$ the band gap, both in eV. For a binary compound $G_gH_h$ with stoichiometry coefficients $g$ and $h$ (e.g. for $Ta_2O_5$: $g = 2$, $h = 5$), the average atomic number $Z$ and thickness per monolayer $a$ are defined in the following way:[4]

$Z = \frac{gZ_g+hZ_h}{g+h}$ with the atomic numbers $Z_g$ and $Z_h$

$a = \sqrt[3]{\frac{M}{\rho N_A(g+h)}}$ with the molecular weight $M$, mass density $\rho$ and the Avogadro constant $N_A$.

In the case of an elemental solid, the equations simplify by $g = 1$ and $h = 0$.

Importantly, this equation should only be used for emission angles $\leq 65°$ and overlayer thicknesses (in our case $Ta_2O_5$) which reduce the substrate intensity (in our case Ta) to maximum 10 % of its original value. Outside of this window, the EALs can no longer be assumed constant, such that a significant error in the analysis can arise.[3]

The parameters used to calculate $L_{\mathrm{Ta_2O_5}}$ and $L_{\mathrm{Ta}}$ and the resulting values for all peak regions investigated can be found in the following two tables S*4* and S*5*.

**Table S*4*:** Material parameters used to calculate $L_{\mathrm{Ta}}$ and $L_{\mathrm{Ta_2O_5}}$.

| | $M$ (u) | $\rho$ (g/cm$^3$) | $a$ (nm) | $E_g$ (eV) | $Z$ |
|---|---|---|---|---|---|
| Ta | 180.95 | 16.7 | 0.26 | 0 | 73 |
| $Ta_2O_5$ | 441.90 | 8.2 | 0.23 | 3.8 | 26.6 |

**Table S*5*:** Values obtained for $L_{\mathrm{Ta}}$ and $L_{\mathrm{Ta_2O_5}}$ for each analyzed peak regions. The kinetic energy $E_{\mathrm{kin}}$ is related to the binding energy $E_{\mathrm{BE}}$ via $E_{\mathrm{kin}} = h\nu - E_{\mathrm{BE}} - \Phi$ with $h\nu = 1486.6$ eV and $\Phi = 4.39$ eV. For the binding energy, the energy of the larger peak of the doublets was used.

| Peak region | $E_{\mathrm{kin,Ta}}$ (eV) | $L_{\mathrm{Ta}}$ (nm) | $E_{\mathrm{kin,Ta_2O_5}}$ (eV) | $L_{\mathrm{Ta_2O_5}}$ (nm) |
|---|---|---|---|---|
| Ta 4f | 1460.6 | 1.53 | 1455.3 | 1.85 |
| Ta 5s | 1412.3 | 1.49 | 1409.0 | 1.80 |
| Ta 4d | 1257.3 | 1.35 | 1251.5 | 1.63 |
| Ta 4p | 1082.8 | 1.21 | 1076.8 | 1.44 |

Additionally to the EALs, equation (S4) also requires the number densities $N_{\mathrm{Ta}^{5+}}$ and $N_{\mathrm{Ta}^0}$. They are defined as the number of $Ta^{5+}$ or $Ta^0$ atoms per unit volume and are given in table S*5*.

**Table S*6***: Values of $N_{\mathrm{Ta}^{5+}}$ and $N_{\mathrm{Ta}^0}$ used in the XPS analysis.

| | $\boldsymbol{N}\ (\mathbf{10^{22}/cm^3})$ |
|---|---|
| Ta | 5.54 |
| $Ta_2O_5$ | 2.23 |

## SI 5. XPS peak fitting

XPS peak fitting was performed in casaXPS version 2.3.26PR1.0. The charge compensation was done by aligning the Fermi edge to a binding energy of 0 eV. Symmetric peaks were fitted using the Gaussian/Lorentzian product formula GL(*p*) where GL(0) is a pure Gaussian and GL(100) a pure Lorentzian. All metallic peaks are asymmetric and were fitted with the Lorentzian asymmetric lineshape LA(*a,b,m*). It is defined as a convolution of a Lorentzian, with a spread to the left and right defined by the parameters *a* and *b*, and a Gaussian with width *m*.

Where possible, backgrounds were subtracted using the Analytic Shirley background. It has the advantage of showing steps at each peak which correspond to the additional inelastic scattering of the corresponding transition, equivalently to the Shirley background. Therefore, the background is physically justified also for regions with several peaks. To take account of broad underlying peaks in the background, the Analytic Shirley background has four cross-section parameters. Three of them define the linear, quadratic and cubic coefficients of a cubic polynomial with which the step height is modified. The fourth parameter is an offset in energy. However, the Analytic Shirley background always ends horizontally towards the lower binding energy side. For regions lying on a tilted background this leads to errors which can significantly change the intensity ratios. Therefore, the backgrounds of the Ta 5s and O 1s regions, which require a tilted background and only span a small energy range, were subtracted using an Offset Shirley background.[5] In the case of the Ta 4p region, the peak region is large and lying on top of a very pronounced peak. Therefore, a Spline Tougaard background was used, which allows for most flexibility to adapt the shape of the background.[5]

For spin-orbit split peak doublets, the ratio of the peak intensities was fixed according to their degeneracy, given by 2J+1: For the 4f-orbitals the intensity ratio is given by $I(4f_{7/2})/I(4f_{5/2}) = 4/3$, for 4d-orbitals it is $I(4d_{5/2})/I(4d_{3/2}) = 3/2$ and for 4p-orbitals it is $I(4p_{3/2})/I(4p_{1/2}) = 2/1$.

In the Ta 4f peak region, $O^{2-}$ 2s (at 22.3 eV in the 2 nm $HfO_2$ samples) and $In^{3+}$ 4d (at 18.9 and 19.7 eV in the 2 nm $HfO_2$ samples) peaks are buried below the other peaks. To minimize errors on the $Ta^0$ 4f and $Ta^{5+}$ 4f peak intensities, the intensities of the $O^{2-}$ 2s and $In^{3+}$ 4d peaks were estimated in the following way: First, the areas of the $O^{2-}$ 1s and $In^{3+}$ 3d peaks were determined in the O 1s and Ta 4p peak region, respectively, where these peaks can be fitted reliably (see tables S*6* and S*7)*. Then, the corresponding areas of the $O^{2-}$ 2s and $In^{3+}$ 4d peaks were calculated using the corrected relative sensitivity factor (RSF) values at 55 eV pass energy for our instrument. The resulting areas were then fixed in the Ta 4f peak region. For simplicity, the $O^{2-}$ 2s peak was thereby only fitted with one asymmetric peak shape instead of the two symmetric peaks corresponding to the carbonate and hydroxide subpeaks.

Tables S*7* and S*8* provide an overview over all fit parameters of the samples with 2 and 3 nm $HfO_2$, respectively. The binding energies and full width half max (FWHM) values were averaged over all emission angles and gate voltages and the average and standard deviation are given in the tables.

**Table S7.** For sample Si/ $SiO_2$/ Ta (15)/ $HfO_2$ (2): Binding energies, line shapes, full width half max (FWHM) and background types for the orbitals in all analyzed peak regions. For the binding energy and FWHM, the average and standard deviation of the values from all angles and all gate voltages are provided (if smaller than 0.1 eV, 0.1 eV is estimated as the uncertainty). Large uncertainties in the FWHM are typically caused by large differences in peak size after different gating steps (the smaller the peak, the smaller its FWHM).

| Peak region | Orbital | Binding energy (eV) | Line shape | FWHM (eV) | Background type |
|---|---|---|---|---|---|
| Ta 4f | $Hf^{4+}$ $4f_{7/2}$ | 17.8 ± 0.1 | GL(40) | 1.2 ± 0.1 | Analytic Shirley |
| | $Hf^{4+}$ $4f_{5/2}$ | 19.5 ± 0.2 | GL(40) | 1.2 ± 0.1 | |
| | $Ta^0$ $4f_{7/2}$ | 21.6 ± 0.1 | LA(1,7,50) | 0.8 ± 0.1 | |
| | $Ta^0$ $4f_{5/2}$ | 23.5 ± 0.1 | LA(1,7,50) | 0.8 ± 0.1 | |
| | $Ta^{5+}$ $4f_{7/2}$ | 26.9 ± 0.2 | GL(40) | 1.3 ± 0.2 | |
| | $Ta^{5+}$ $4f_{5/2}$ | 28.8 ± 0.1 | GL(40) | 1.3 ± 0.2 | |
| | $O^{2-}$ 2s | 22.3 ± 0.2 | LA(1,7,150) | 5.5 ± 0.1 | |
| | $In^{3+}$ $4d_{5/2}$ | 18.9 ± 0.2 | GL(50) | 0.7 ± 0.1 | |
| | $In^{3+}$ $4d_{3/2}$ | 19.7 ± 0.2 | GL(50) | 0.7 ± 0.1 | |
| O 1s | $O^{2-}$ 1s | 531.1 ± 0.2 | GL(40) | 1.4 ± 0.1 | Offset Shirley |
| | O 1s (O-C) | 532.6 ± 0.2 | GL(40) | 2.0 ± 0.1 | |
| Ta 5s | $Hf^{4+}$ 5s | 65.9 ± 0.2 | GL(75) | 4.6 ± 0.9 | Offset Shirley |
| | $Ta^0$ 5s | 69.9 ± 0.2 | LA(1.4,6,80) | 4.2 ± 1.2 | |
| | $Ta^{5+}$ 5s | 73.2 ± 0.2 | GL(95) | 3.9 ± 1.0 | |
| Ta 4d | $Hf^{4+}$ $4d_{5/2}$ | 214.0 ± 0.2 | GL(80) | 4.1 ± 0.1 | Analytic Shirley |
| | $Hf^{4+}$ $4d_{3/2}$ | 224.9 ± 0.2 | GL(80) | 4.6 ± 0.3 | |
| | $Ta^0$ $4d_{5/2}$ | 224.9 ± 0.2 | LA(1.3,2,50) | 3.6 ± 0.9 | |
| | $Ta^0$ $4d_{3/2}$ | 238.0 ± 0.2 | LA(1.3,2,50) | 4.0 ± 0.8 | |
| | $Ta^{5+}$ $4d_{5/2}$ | 230.7 ± 0.3 | GL(95) | 3.8 ± 0.6 | |
| | $Ta^{5+}$ $4d_{3/2}$ | 242.5 ± 0.1 | GL(80) | 3.8 ± 0.8 | |
| Ta 4p | $Hf^{4+}$ $4p_{3/2}$ | 382.7 ± 0.1 | GL(85) | 5.0 ± 0.1 | Spline Tougaard |
| | $Hf^{4+}$ $4p_{1/2}$ | 439.9 ± 0.1 | GL(95) | 7.3 ± 0.2 | |
| | $Hf^{4+}$ (sat) | 396.4 ± 0.2 | GL(30) | 5.8 ± 0.7 | |
| | $Hf^{4+}$ (sat) | 453.1 ± 0.5 | GL(30) | 6.0 ± 0.7 | |
| | $Ta^0$ $4p_{3/2}$ | 399.4 ± 0.2 | LA(1,7,150) | 4.7 ± 1.2 | |
| | $Ta^0$ $4p_{1/2}$ | 458.9 ± 1.1 | LA(1,7,150) | 7.7 ± 2.6 | |
| | $Ta^0$ (sat) | 421.1 ± 0.7 | GL(30) | 9.4 ± 1.6 | |
| | $Ta^0$ (sat) | 484.6 ± 1.7 | GL(30) | 8.7 ± 2.8 | |
| | $Ta^{5+}$ $4p_{3/2}$ | 405.4 ± 0.3 | GL(77) | 6.0 ± 1.1 | |
| | $Ta^{5+}$ $4p_{1/2}$ | 466.4 ± 0.4 | GL(77) | 8.8 ± 1.9 | |
| | $In^{3+}$ $3d_{5/2}$ | 445.9 ± 0.1 | GL(50) | 1.5 ± 0.1 | |
| | $In^{3+}$ $3d_{3/2}$ | 453.4 ± 0.2 | GL(50) | 1.5 ± 0.1 | |

**Table S*8*.** For sample Si/ $SiO_2$/ Ta (15)/ $HfO_2$ (3): Binding energies, line shapes, full width half max (FWHM) and background types for the orbitals in all analyzed peak regions, equivalently to Table S*7* for the sample with 2 nm $HfO_2$.

| Peak region | Orbital | Binding energy (eV) | Line shape | FWHM (eV) | Background type |
|---|---|---|---|---|---|
| Ta 4f | $Hf^{4+}$ $4f_{7/2}$ | 18.1 ± 0.1 | GL(30) | 1.2 ± 0.1 | Analytic Shirley |
| | $Hf^{4+}$ $4f_{5/2}$ | 19.8 ± 0.1 | GL(40) | 1.2 ± 0.2 | |
| | $Ta^{0}$ $4f_{7/2}$ | 21.7 ± 0.2 | LA(1,7,50) | 0.7 ± 0.1 | |
| | $Ta^{0}$ $4f_{5/2}$ | 23.6 ± 0.2 | LA(1,7,50) | 0.7 ± 0.1 | |
| | $Ta^{5+}$ $4f_{7/2}$ | 27.2 ± 0.1 | GL(40) | 1.9 ± 0.7 | |
| | $Ta^{5+}$ $4f_{5/2}$ | 29.0 ± 0.1 | GL(40) | 1.9 ± 0.7 | |
| | $O^{2-}$ 2s | 22.1 ± 0.1 | LA(1,7,150) | 5.4 ± 0.3 | |
| | $In^{3+}$ $4d_{5/2}$ | 19.8 ± 0.1 | GL(50) | 1.0 ± 0.1 | |
| | $In^{3+}$ $4d_{3/2}$ | 20.6 ± 0.1 | GL(50) | 1.0 ± 0.1 | |
| O 1s | $O^{2-}$ 1s | 531.4 ± 0.1 | GL(40) | 1.4 ± 0.1 | Offset Shirley |
| | O 1s (O-C) | 532.9 ± 0.1 | GL(40) | 2.0 ± 0.1 | |
| Ta 5s | $Hf^{4+}$ 5s | 66.3 ± 0.2 | GL(75) | 4.8 ± 0.6 | Offset Shirley |
| | $Ta^{0}$ 5s | 70.3 ± 0.1 | LA(1.4,6,80) | 3.2 ± 0.9 | |
| | $Ta^{5+}$ 5s | 73.2 ± 0.2 | GL(95) | 2.8 ± 1.4 | |
| Ta 4d | $Hf^{4+}$ $4d_{5/2}$ | 214.4 ± 0.1 | GL(80) | 4.2 ± 0.1 | Analytic Shirley |
| | $Hf^{4+}$ $4d_{3/2}$ | 225.0 ± 0.1 | GL(80) | 4.9 ± 0.4 | |
| | $Ta^{0}$ $4d_{5/2}$ | 225.4 ± 0.2 | LA(1.3,2,50) | 3.0 ± 0.5 | |
| | $Ta^{0}$ $4d_{3/2}$ | 238.1 ± 0.2 | LA(1.3,2,50) | 4.1 ± 0.7 | |
| | $Ta^{5+}$ $4d_{5/2}$ | 230.9 ± 0.3 | GL(95) | 4.9 ± 0.6 | |
| | $Ta^{5+}$ $4d_{3/2}$ | 242.7 ± 0.3 | GL(85) | 3.3 ± 1.2 | |
| Ta 4p | $Hf^{4+}$ $4p_{3/2}$ | 383.0 ± 0.1 | GL(85) | 4.9 ± 0.1 | Spline Tougaard |
| | $Hf^{4+}$ $4p_{1/2}$ | 440.3 ± 0.1 | GL(95) | 7.5 ± 0.2 | |
| | $Hf^{4+}$ (sat) | 396.7 ± 0.1 | GL(30) | 6.0 ± 0.8 | |
| | $Hf^{4+}$ (sat) | 454.2 ± 0.4 | GL(30) | 5.5 ± 0.8 | |
| | $Ta^{0}$ $4p_{3/2}$ | 399.6 ± 0.3 | LA(1,7,150) | 5.1 ± 1.0 | |
| | $Ta^{0}$ $4p_{1/2}$ | 459.6 ± 0.6 | LA(1,7,95) | 7.6 ± 2.5 | |
| | $Ta^{0}$ (sat) | 420.7 ± 0.2 | GL(30) | 7.6 ± 1.4 | |
| | $Ta^{0}$ (sat) | 487.1 ± 0.9 | GL(30) | 3.6 ± 1.6 | |
| | $Ta^{5+}$ $4p_{3/2}$ | 405.6 ± 0.3 | GL(77) | 7.1 ± 1.2 | |
| | $Ta^{5+}$ $4p_{1/2}$ | 465.3 ± 0.8 | GL(77) | 7.9 ± 2.7 | |
| | $In^{3+}$ $3d_{5/2}$ | 446.2 ± 0.1 | GL(50) | 1.4 ± 0.1 | |
| | $In^{3+}$ $3d_{3/2}$ | 453.7 ± 0.1 | GL(50) | 1.4 ± 0.1 | |

## SI 6. Overview over the XPS fits for all peak regions at all angles and gate voltages

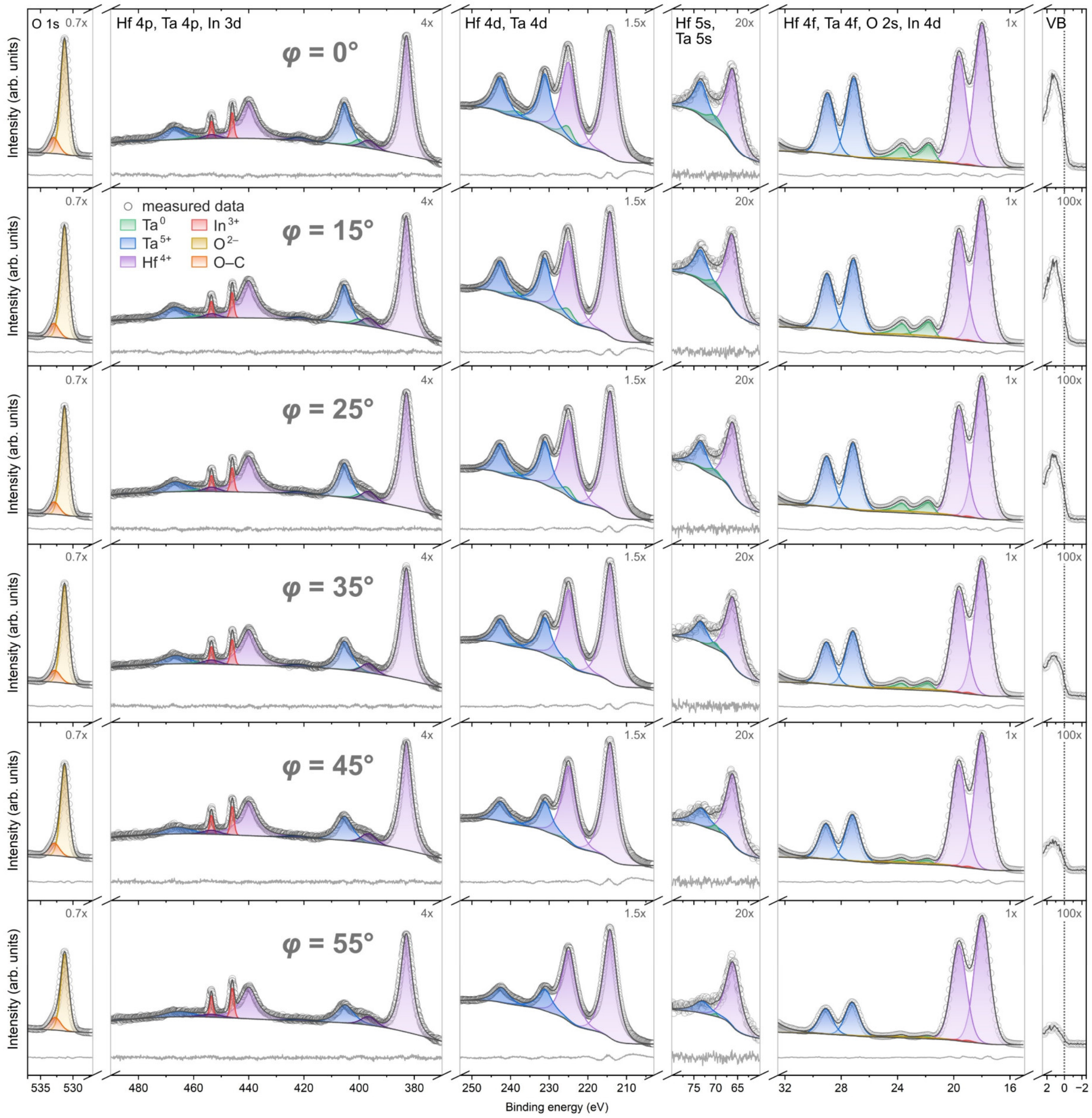

**Figure S2:** Overview over the XPS spectra for the sample with 2 nm $HfO_2$ after -3 V gating for all analyzed peak regions at all angles.

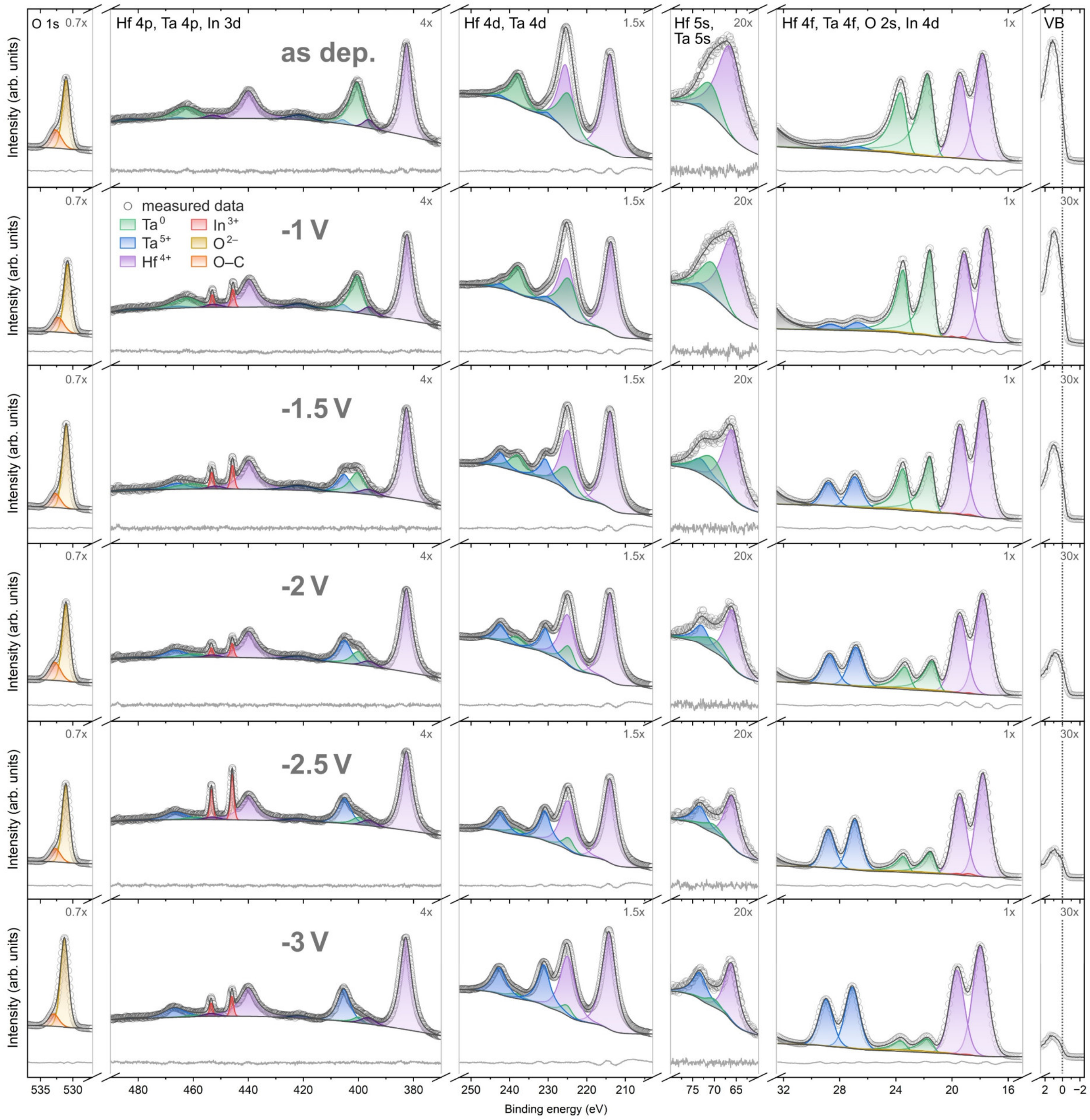

**Figure S3:** Overview over the XPS spectra for the sample with 2 nm $HfO_2$ at 0° emission angle for all analyzed peak regions at all gating steps.

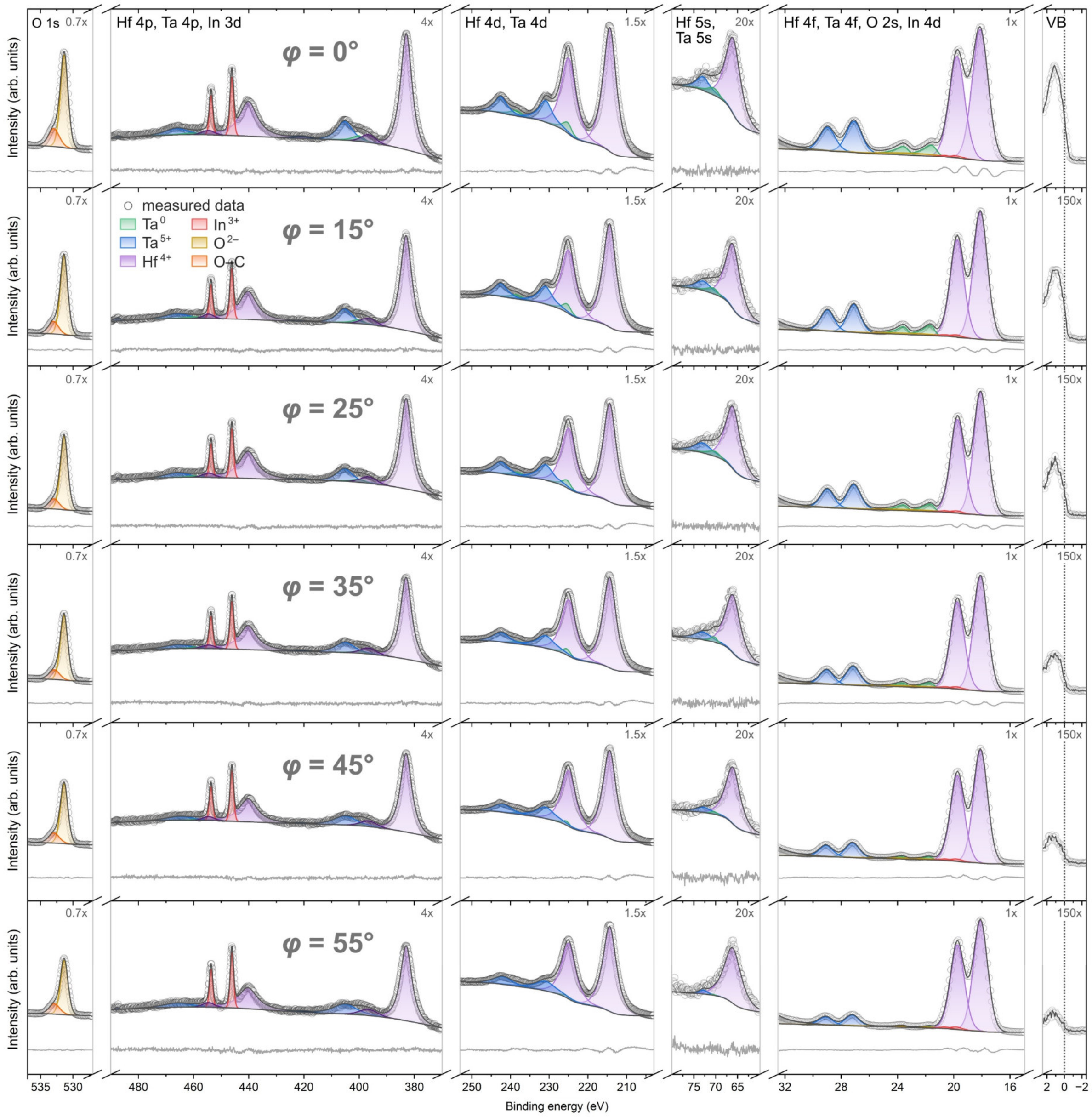

**Figure S4:** Overview over the XPS spectra for the sample with 3 nm $HfO_2$ after -3 V gating for all analyzed peak regions at all angles.

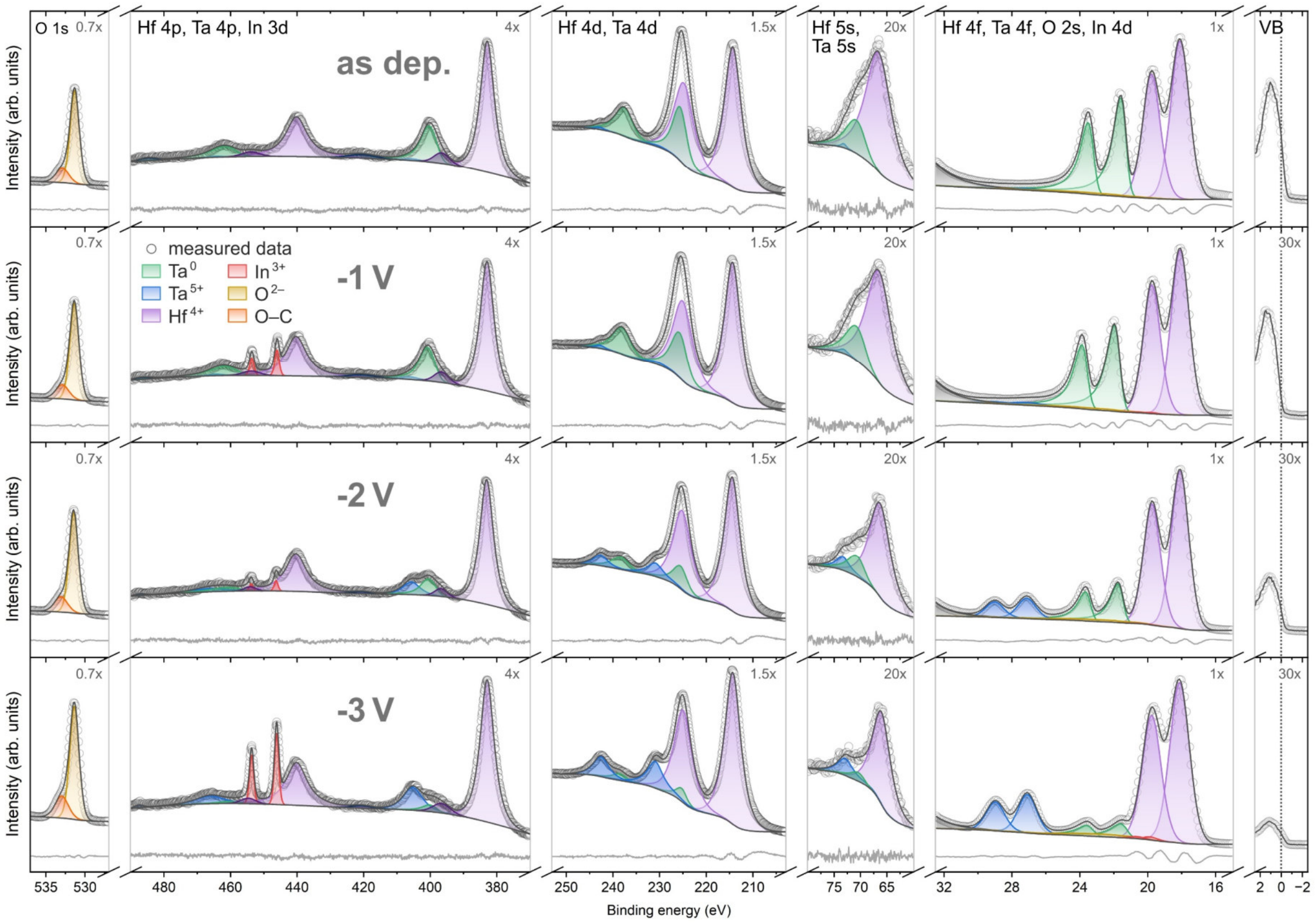

**Figure S5:** Overview over the XPS spectra for the sample with 3 nm $HfO_2$ at 0° emission angle for all analyzed peak regions at all gating steps.

## SI 7. Calculation of the scattering length density (SLD)

In Fig. 4 a of the manuscript, we show a SLD profile calculated from the XPS results for a direct comparison to the SLD obtained from XRR experiments. The SLD, as given in XRR, is the measure of how strongly a material scatters incoming X-rays. It is given by the electron density $\rho_e$ multiplied by the electron radius $r_e$. To calculate the electron density, the scattering factors $f_0$ (atomic form factor, approximately equal to the atomic number $Z$ for small angles $\theta$ towards the sample surface) and $f'$ (dispersion correction) have to be considered for each element. The sum of the two $f_1(E) = (f_0 + f')(E)$ as a function of the energy of the incoming X-rays is tabulated on the website 'The atomic scattering factors' provided by the Berkely lab.[6] The first column gives the energy in eV, the second column gives the scattering factors $f_1 = (f_0 + f')$ and the third column gives the anomalous dispersion factor $f''$ which takes absorption into account and is not required here.

The electron density can be calculated from the product of the scattering factors $(f_0 + f')$ and the number density $N$ of the respective element. For compounds with several elements, the SLD as a function of the energy $E$ is therefore given as

$$\mathrm{SLD}(E) = r_e\, N_{\mathrm{FU}} \sum_{i \in \mathrm{FU}} (f_0 + f')_i(E) \cdot m_i \tag{S6}$$

with the number density $N_{\mathrm{FU}}$ of the formula unit and the stoichiometry coefficients $m_i$ of each element in the formula unit.

To calculate the SLD from the XPS depth profiles, the number densities $N_{\mathrm{Ta}^0}$ and $N_{\mathrm{Ta}^{5+}}$ obtained from XPS as a function of the $z$-position were used (see table S*6*). The number density of $Ta_2O_5$ is obtained from $N_{\mathrm{Ta}_2O_5} = N_{\mathrm{Ta}^{5+}}/2$. Table S*9* provides the parameters for the calculation of the SLDs of pure Ta and $Ta_2O_5$. The SLD profile at the $Ta_2O_5$/Ta interface is obtained by including the z-dependence of the number densities and including both Ta and $Ta_2O_5$ in the sum of equation S6.

**Table S*9*:** Parameters for the calculation of the SLDs of pure Ta and $Ta_2O_5$ at the energy of Cu K$\alpha$ radiation (8.05 keV) as used in XRR. The values of $(f_0 + f')_i$ were obtained from the table from the Berkely lab at 8047.42 eV for Ta and 8048.79 eV for O.

| | $N_{\mathrm{FU}}$ $(10^{22}/\mathrm{cm}^3)$ | $m_{\mathrm{Ta}}$ | $(f_0 + f')_{\mathrm{Ta}}$ | $m_{\mathrm{O}}$ | $(f_0 + f')_{\mathrm{O}}$ | SLD $(r_e/\mathrm{Å}^3)$ |
|---|---|---|---|---|---|---|
| Ta | 5.54 | 1 | 67.51 | 0 | 8.05 | 3.74 |
| $Ta_2O_5$ | 1.12 | 2 | 67.51 | 5 | 8.05 | 1.96 |